\documentclass[a4paper]{jpconf}
\usepackage{graphicx}
\begin{document}
\title{Geodesic behaviour around cosmological milestones}

\author{L. Fern\'andez-Jambrina$^1$ and R. Lazkoz$^2$}

\address{$^1$ Matem\'atica Aplicada, E.T.S.I. Navales, Universidad
Polit\'ecnica de Madrid, Arco de la Victoria s/n,  E-28040 Madrid, Spain}

\address{$^2$ F\'\i sica Te\'orica, Facultad de Ciencia y Tecnolog\'\i
a, Universidad del Pa\'\i s Vasco,\\ Apdo.  644, E-48080 Bilbao, Spain}

\ead{leonardo.fernandez@upm.es, ruth.lazkoz@ehu.es}

\begin{abstract}
In this paper we provide a thorough classification of
Friedman-Lema\^{i}tre-Robertson-Walker (FLRW) cosmological models in
terms of the strong or weak character of their singularities according
to the usual definitions.  The classification refers to a generalised
Puiseux power expansion of the scale factor of the model around a
singular event.
\end{abstract}

\section{Introduction}
In the last few years the number of discussed matter contents for the 
different ages of our universe has increased. Besides the original dust, 
radiation and cosmological constant terms, researchers have come to 
include other contributions to the right-hand side of Einstein 
equations, being quintessence and phantom energy some of the most 
usual. The consideration of these new terms has been motivated by the attempts 
of explaining the experimentally inferred accelerated expansion of 
the universe \cite{de}. 

The inclusion of these new matter contents, which do not satisfy some
of the classical energy conditions, has as a consequence the
appearance of new types of singularities which did not come up in
former models, such as big rip \cite{caldwell} and sudden
singularities \cite{barrow}, and other non-singular features, such as
bounces or extremality events.  All of these have been comprised under
the name ``cosmological milestones'' by Catto\"en and Visser
\cite{visser} together with the classical big bag and big crunch
singularities.

In that paper the authors perform a classification of FLRW
cosmological models in terms of the first coefficients of a
generalised Puiseux expansion of the scale factor in time coordinate
around one of these cosmological milestones.  These coefficients are
used to determine violations of energy conditions and appearance of
polynomial scalar curvature singularities and derivative curvature
singularities. Generically the classification depends on just the 
first three exponents of the Puiseux expansion.

What we would like to do now is to complete the classification by
considering other definitions of singularities.  Geodesic
incompleteness is commonly accepted as an indicator of the existence
of a singularity in a space-time \cite{HE} and may happen even in cases where
there is no polynomial scalar curvature singularity.

Furthermore, it has been shown that even in the cases where a causal 
geodesic is incomplete, this does not mean that finite objects are 
necessarily crushed on approaching the singularity. These are 
considered weak singularities. In our classification we take into 
account this fact with the most common definitions of strong 
singularities.

This involves calculation of causal geodesics in the corresponding 
space-times. This topic is reviewed in section \ref{sec2}. In section 
\ref{sec3} 
geodesic equations are solved for the power expansion of the scale 
factor and the differentiability of the geodesics is analysed. 
Finally, in section \ref{sec4} the strength of the singularities, if any, is 
discussed in relation to the values of the exponents of the power 
expansion. The conclusions are summarized in section \ref{sec5}. More 
details about this issue may be found in \cite{puiseux}.

\section{Geodesic equations\label{sec2}}
Generally calculation of geodesics in a space-time is a cumbersome 
task, since it requires solving a system of four ordinary quasilinear 
differential equations. Geodesics are parametrised by their proper 
time,
\begin{equation}
d\tau^2=-g_{ij}dx^idx^j,
\end{equation}
where $g_{ij}$ are the components of the metric tensor of the 
space-time in the coordinate chart provided by 
$\{x_{0},x_{1},x_{2},x_{3}\}$. Proper time is defined up to a change 
of scale and origin, $\tilde\tau=a\tau+b$, and hence it is also 
called affine parameter. We denote by 
a dot derivatives with respect to proper time.

A geodesic is said to be complete if it can defined for all values of
$\tau$.  On the contrary, it is said to be incomplete in the past
(future) if it can be extended just to a value $\tau_{0}$ instead of
$-\infty$ ($\infty$).

Geodesics are defined as the curves $\Gamma$ on the space-time for which the 
length functional,
\begin{equation}
L[\Gamma]=\int_{s_{0}}^{s_{1}} ds,\quad ds^2=-d\tau^2=g_{ij}dx^idx^j,
\end{equation}
has a extremum. 

The corresponding Euler-Lagrange equations,
\begin{equation}
\ddot x^{i}+\Gamma^i_{jk}\dot x^j\dot x^k=0,\end{equation}
may be written in terms of the Christoffel symbols for the metric 
tensor,
\begin{equation}
\Gamma^{i}_{jk}=\frac{1}{2}g^{il}\left\{g_{lj,k}+g_{lk,j}-g_{jk,l}\right\},
\end{equation}
plus an additional equation,
\begin{equation}\label{delta}
\delta=-g_{ij}\dot x^{i}\dot x^{j}, 
\end{equation}
which simply states that we are using proper time as parameter. The 
constant $\delta$ takes value one for timelike, zero for lightlike 
and minus one for spacelike geodesics.

We use spherical coordinates, $\{t,r,\theta,\phi\}$, with the 
usual ranges and $t$ is coordinate time. Geodesics are therefore  
described providing 
$\left(t(\tau),r(\tau),\theta(\tau),\phi(\tau)\right)$. 

In the case of FLRW cosmological models,
\begin{eqnarray}
&&ds^2=-dt^2+a^2(t)\left\{f^2(r)dr^2+ r^2\left(d\theta^2+\sin^2\theta
d\phi^2\right)\right\}\nonumber\\
&&f^2(r)=\frac{1}{1-kr^2},\quad k=0,\pm1,\label{metric}\end{eqnarray}
the large number of isometries allows us a quick integration of 
geodesic equations.

Since the space-time is homogeneous and isotropic, geodesics are 
straight lines and hence we may restrict the discussion to a 
plane $\theta=\pi/2$, $\phi=\mathrm{const}$, choosing as origin of 
coordinates one of the points of the geodesic. 

Furthermore, changing the radial coordinate to 
\begin{equation}
R=\left\{\begin{array}{ll}\mathrm{arcsinh}\, r & k=-1  \\r & k=0  \\
\arcsin r & k=1\end{array}\right.,\quad ds^2=-dt^2+a^2(t)
\left\{dR^2+ \begin{array}{c}\sinh^2R\\R^2\\\sin^2R\end{array}\left(d\theta^2+\sin^2\theta
d\phi^2\right)\right\},
\end{equation}
it is easy to check that $\partial_{R}=\partial_{r}/f(r)$ is another 
generator of isometries and therefore
\begin{equation}
P=a^2(t)f(r)\dot r
\end{equation}
is conserved along geodesics. 

We are left then with just one equation,
\begin{equation}
\dot t^2=\delta+\frac{P^2}{a^2(t)}
\end{equation}
and the quadrature
\begin{equation}
\dot r=\frac{P}{a^2(t)f(r)},
\end{equation}
which may be integrated after solving the equation in $t$ and 
therefore need not be considered here.

At this point it is clear that all information about the geodesics is 
encoded in the scale factor $a(t)$. 

We consider just future-pointing geodesics, $\dot t>0$.

Without losing much generality, we assume that the expansion,
\begin{equation}\label{puiseux}
    a(t)=c_{0}|t-t_{0}|^{\eta_{0}}+c_{1}|t-t_{1}|^{\eta_{1}}+\cdots,
\end{equation}
where the exponents $\eta_{i}$ are real and ordered,
\[\eta_{0}<\eta_{1}<\cdots\]
is valid close to a cosmological milestone at $t_{0}$. The 
coefficient $c_{0}$ must be positive in order to have a positive 
scale factor.

We consider just singularities 
in the past, $t>t_{0}$, in order to avoid signs and absolute values. 
Since equations are time-reversal symmetric, no information is lost 
with this restriction.

At lowest order, $\eta_{0}$, in the flat universe case, $k=0$, the
model behaves like a perfect fluid of density $\rho$ and pressure $p$
with a linear equation of state,
\begin{equation}
p=w\rho,\quad w=-1+\frac{2}{3} \eta_{0}.
\end{equation}

Also at lowest order, three different behaviours are possible for the
scale factor at $t_{0}$: zero, finite and divergent:

\begin{itemize}
    \item  $\eta_{0}>0$: the scale factor vanishes at $t_{0}$ and
    generically we have a big bang or big crunch singularity.

    \item  $\eta_{0}=0$: the scale factor is finite at $t_{0}$.
    If $a(t)$ is analytical, the event at $t_{0}$ is regular.
    Otherwise a sudden singularity comes up 
    \cite{barrow,barrowfollow}.

    \item  $\eta_{0}<0$: the scale factor diverges at $t_{0}$ and a
    big rip singularity comes up.
\end{itemize}

Since completeness of just causal geodesics is required for the 
analysis of singularities, we focus only on lightlike and timelike 
geodesics.

\section{Geodesic completeness of causal geodesics\label{sec3}}

Lightlike geodesic equations are straightforwardly integrated,
\begin{equation}
a(t)\dot t=P \Rightarrow \int_{t_{0}}^t a(t')\,dt'=P(\tau-\tau_{0}).
\end{equation}

At lowest order,
\begin{equation}a(t)\simeq c_{0}|t-t_{0}|^{\eta_{0}}\Rightarrow
t\simeq t_{0}+\left\{\begin{array}{ll}\left\{\frac{(1+\eta_{0}) P}{c_{0}}\right\}^{1/(1+\eta_{0})}
(\tau-\tau_{0})^{1/(1+\eta_{0})} &\eta_{0}\neq -1\\ \\ Ce^{P\tau/c_{0}}
&\eta_{0}=-1\end{array}\right..
\end{equation}

Since generically $t$ behaves as a power $1/(1+\eta_{0})$ of proper 
time, different levels of regularity appear depending on the value of 
$\eta_{0}$. 

It is worth mentioning that for $\eta_{0}$ lower or equal than minus one, 
lightlike geodesics do not reach the cosmological milestone at 
$t_{0}$, since it would take them an infinite proper time to reach 
it. They therefore do not see the singularity. This limiting case, 
which corresponds to a model with $w=-5/3$ when the universe is flat,  
has been named superphantom and considered in \cite{dabrowski}.

Results on the differentiability of lightlike geodesics at $\eta_{0}$
are consigned in table \ref{table1}.

\begin{table}
\caption{\label{table1}Derivatives of lightlike geodesics at $t_{0}$.}
\begin{center}
\begin{tabular}{llllll}
    \br
   $\eta_{0}$ & $\eta_{1}$ & $\dot t$ & $\ddot t$ & $\stackrel{\dots}{t}$ &
$t^{n)}$  \\
    \mr
   $(0,\infty) $ & $(\eta_{0},\infty)$ & $\infty$ & $\infty$ &
$\infty$ & $\infty$  \\
  $0$ & $(0,1)$ & finite & $\infty$ & $\infty$ & $\infty$  \\
  & $(1,2)$ & finite & finite & $\infty$ & $\infty$  \\
 & $(2,3)$ & finite & finite & finite & $\infty$  \\
  $(-1/2,0)$ & $(\eta_{0},\infty)$ & finite & $\infty$ & $\infty$ &
$\infty$  \\
 $-1/2$ & $(-1/2,0)$ & finite & finite & $\infty$ & $\infty$  \\
 & $(0,1/2)$ & finite & finite & finite & $\infty$  \\
$(-2/3,-1/2)$ & $(\eta_{0},\infty)$ & finite & finite & $\infty$ &
$\infty$  \\
$-2/3$ & $(-2/3,-1/3)$ & finite & finite & finite & $\infty$  \\
    $\big(\frac{1-n}{n},\frac{2-n}{n-1}\big)$ & $(\eta_{0},\infty)$ &
finite &
   finite & finite & $\infty$  \\
    $(-\infty, -1]$  & $(\eta_{0},\infty)$ & / & / & / & / \\
    \br
\end{tabular}
\end{center}
\end{table}

For the values of $\eta_{0}$ for which the class of differentiability 
increases by one, it is necessary to consider further terms of the 
Puiseux expansion.

We notice that, as $\eta_{0}$ decreases, the class of 
differentiability increases.

Timelike geodesics may be analysed similarly, though in this case 
geodesic equations cannot be solved analytically.

At lowest order,
\begin{equation}\label{ttime}
\dot t=\sqrt{1+\frac{P^2}{a^2}}\simeq
\sqrt{1+\frac{P^2}{c_{0}^2}(t-t_{0})^{-2\eta_{0}}}
,\end{equation}
the geodesic is singular at $t=t_{0}$ for
$\eta_{0}>0$, since $\dot t$ blows up.

However, the derivative is well defined for negative (big rip)
$\eta_{0}$.  Near $t_{0}$ we may write
\begin{equation}
\dot t \simeq 1+\frac{P^2}{2c_{0}^2}(t-t_{0})^{-2\eta_{0}},\end{equation}
and similar expressions for higher derivatives,
\begin{equation}
t^{n)}\sim(t-t_{0})^{-2\eta_{0}-n+1}.
\end{equation}

Again, as it is shown in table \ref{table2}, the class of differentiability of 
timelike geodesics increases as $\eta_{0}$ decreases. However, every 
timelike geodesic reaches the cosmological milestone at $t_{0}$ and 
there are no curves with all finite derivatives there.

\begin{table}
\caption{\label{table2}Derivatives of timelike geodesics at $t_{0}$.}
\begin{center}
\begin{tabular}{llllll}
    \br
   $\eta_{0}$ & $\eta_{1}$ & $\dot t$ & $\ddot t$ & $\stackrel{\dots}{t}$ &
$t^{n)}$  \\
    \mr
   $(0,\infty) $ & $(\eta_{0},\infty)$ & $\infty$ & $\infty$ &
$\infty$ & $\infty$  \\
  $0$ & $(0,1)$ & finite & $\infty$ & $\infty$ & $\infty$  \\
  & $(1,2)$ & finite & finite & $\infty$ & $\infty$  \\
 & $(2,3)$ & finite & finite & finite & $\infty$  \\
  $(-1/2,0)$ & $(\eta_{0},\infty)$ & finite & $\infty$ & $\infty$ &
$\infty$  \\
 $-1/2$ & $(-1/2,1/2)$ & finite & finite & $\infty$ & $\infty$  \\
 & $(1/2,3/2)$ & finite & finite & finite & $\infty$  \\
$(-1,-1/2)$ & $(\eta_{0},\infty)$ & finite & finite & $\infty$ &
$\infty$  \\
$-1$ & $(-1,0)$ & finite & finite & finite & $\infty$  \\
    $\big(\frac{1-n}{2},\frac{2-n}{2}\big)$ & $(\eta_{0},\infty)$ &
finite &
   finite & finite & $\infty$ \\
\br
\end{tabular}
\end{center}
\end{table}

\section{Strength of singularities\label{sec4}}

In the previous section we have shown that qualitatively the strength 
of singularities at $t_{0}$ decreases with $\eta_{0}$, since the 
class of differentiability increases. We proceed now to check this 
qualitative statement with the most usual definitions of strong 
singularities.

Roughly speaking, a singularity is considered strong if tidal forces 
are capable of disrupting a finite object falling into it \cite{ellis}. 

This concept has been developed in several ways. For Tipler 
\cite{tipler} the finite volume is spanned by three Jacobi fields 
that form an orthonormal basis with the velocity $u$ of the geodesic. 
The singularity if strong if the volume tends to zero on approaching 
the singularity.

Another definition is due to Kr\'olak \cite{krolak}, for which the singularity is
strong if the derivative of the volume is negative close to the
singularity.  Obviously, if a singularity is strong according to
Tipler's definition, it is strong according to Kr\'olak's, but not 
conversely.

These definitions are meant to be used for gravitational collapse and 
therefore do not consider the possibility of big rip singularities. 
But these may be included in the framework just reversing a sign.

Both definitions have been written in an amenable form by Clarke and 
Kr\'olak \cite{clarke} in terms of integrals of Riemann components 
along the geodesics. 

In our case the situation is even much simpler since the space-time is 
conformally flat and the Weyl tensor vanishes.

For instance, a lightlike geodesic meets a
strong singularity, according to Tipler's definition, at proper time
$\tau_{0}$ if and only if 
\begin{equation}\label{suftipler}
 \int_{0}^{\tau}d\tau'\int_{0}^{\tau'}d\tau''R_{ij}u^{i}u^j
\end{equation}
diverges as $\tau$ tends to $\tau_{0}$.

And a lightlike geodesic meets a strong singularity 
at proper time $\tau_{0}$ if and only if
\begin{equation}\label{sufkrolak}
 \int_{0}^{\tau}d\tau'R_{ij}u^{i}u^j
 \end{equation}
diverges as $\tau$ tends to $\tau_{0}$.

In our case, the velocity of the
geodesic is
\begin{equation}
R_{ij}u^iu^j=2P^2\left(\frac{a'^2+k}{a^4}-\frac{a''}{a^3}\right)\simeq
\frac{2P^2\eta_{0}}{c_{0}^2|t-t_{0}|^{2(\eta_{0}+1)}}
+\frac{2kP^2}{c_{0}^4|t-{t_{0}}|^{4\eta_{0}}}
+\cdots,\end{equation}
and at lowest order there are two cases, depending on whether the 
curvature term dominates over the first term.

The results according to both definitions are summarized in table 
\ref{table3}.

\begin{table}
\caption{\label{table3}Degree of singularity of null geodesics around $t_{0}$.}
\begin{center}
   \begin{tabular}{llllll}
    \br
    ${\eta_{0}}$ & ${\eta_{1}}$ & ${k}$ & $c_{0}$ &\textbf{Tipler} &
    \textbf{Kr\'olak}  \\
    \mr
    $(-\infty,-1]$ & $(\eta_{0},\infty)$ &  $0,\pm 1$ & $(0,\infty)$& Regular & Regular  \\
	$(-1,0)$ &  &  & & Strong &
Strong  \\
    $0$ & $(0,1)$ &  && Weak 
    & Strong  \\
	 & $[1,\infty)$ &   && Weak 
	 & Weak 
	  \\
    $(0,1)$ & $(\eta_{0},\infty)$ &   && Strong & Strong  \\
    $1$ & $(1,\infty)$ & $0,1$  && Strong & Strong  \\
     & $(1,\infty)$ & $-1$ &$(0,1)\cup(1,\infty)$ & Strong & Strong  \\
     & $(1,3)$ &  & 1 &Weak
     & Strong  \\
     & $[3,\infty)$ &   &  &Weak 
     & Weak 
     \\
    $(1,\infty)$ & $(\eta_{0},\infty)$ & $0,\pm 1$  &$(0,\infty)$&
Strong & Strong  \\
    \br
\end{tabular}
\end{center}
\end{table}

As we see, besides the models with $\eta_{0}\le -1$, for which no 
lightlike geodesic reaches the cosmological milestone at $t_{0}$, 
there are only two cases without strong singularities: some of the 
cases with $\eta_{0}=0$, which are named sudden singularities, some 
of which had already been studied in \cite{flrw}; and some of the 
cases with $\eta_{0}=1$, $c_{0}=1$, $k=-1$, which are at first order 
Milne universe, which is Minkowski empty space in other coordinates.

Again, there are limiting cases which require resorting to further 
terms in the expansion.

As it was pointed out, it is explicitly checked that Kr\'olak's 
definition includes more cases than Tipler's.

The analysis of the strength of singularities of timelike geodesics 
is somewhat more involved, since there are no both necessary and 
sufficient conditions for the appearance of strong singularities:

According to Tipler's definition, a timelike geodesic meets a
strong singularity at proper time $\tau_{0}$ if
\begin{equation}\label{suftipler1}
 \int_{0}^{\tau}d\tau'\int_{0}^{\tau'}d\tau''R_{ij}u^{i}u^j
 \end{equation}
diverges as $\tau$ tends to $\tau_{0}$.

With Kr\'olak's definition, a timelike geodesic meets a
strong singularity at proper time $\tau_{0}$ if
\begin{equation}\label{sufkrolak1}
 \int_{0}^{\tau}d\tau'R_{ij}u^{i}u^j
 \end{equation}
diverges as $\tau$ tends to $\tau_{0}$.

Necessary conditions are slightly different.  With Tipler's definition
\cite{clarke}, if a timelike geodesic meets a strong
singularity, then 
\begin{equation} I^i_{j}(\tau)=\int_{0}^\tau d\tau'\int_{0}^{\tau'}
    d\tau''\left|R^i_{kjl}u^ku^l\right|,\end{equation}
diverges as $\tau$ tends to $\tau_{0}$ for some $i$, $j$, where the
components are referred to a parallely transported orthonormal frame.

With Kr\'olak's definition, if a timelike geodesic meets a strong
singularity, then 
\begin{equation} I^i_{j}(\tau)=\int_{0}^\tau d\tau'
    \left|R^i_{kjl}u^ku^l\right|,\end{equation}
diverges as $\tau$ tends to $\tau_{0}$ for some $i$, $j$.

Fortunately, this set of sufficient and necessary conditions is 
accurate enough to allow a thorough classification of the 
singularities of timelike geodesics in FLRW models. 

There are two sets of timelike geodesics with different behaviour:

Timelike geodesics with $P=0$, for which the time coordinate is 
essentially proper time, 
\begin{equation}t-t_{0}=\tau-\tau_{0}\;,\quad r=r_{0},\end{equation}
form the congruence of fluid worldlines, 
since the coordinates are comoving for the perfect fluid, 
$u=\partial_{t}$, and do not suffer therefore any problems of 
differentiability. On applying the conditions for the appearance of 
strong singularities, we reach the results of table \ref{table4}.

\begin{table}
    \caption{\label{table4}
     Degree of singularity of  the fluid congruence  of timelike
    geodesics around $t_{0}$.}\begin{center}
    \begin{tabular}{llll}
    \br
    ${\eta_{0}}$ & ${\eta_{1}}$  & \textbf{Tipler} &
    \textbf{Kr\'olak}  \\
    \mr
    $(-\infty,0)$ & $(\eta_{0},\infty)$ & Strong & Strong  \\
	$0$ & $(0,1)$  & Weak & Strong  \\
	 & $[1,\infty)$  & Weak & Complete  \\
    $(0,1)$ & $(\eta_{0},\infty)$ & Strong & Strong  \\
    $1$ & $(1,2]$ &  Weak & Strong  \\
     & $(2,\infty)$ &  Weak & Weak  \\
    $(1,\infty)$ & $(\eta_{0},\infty)$ &  Strong & Strong  \\
    \br
\end{tabular}
\end{center}
\end{table}

Along timelike geodesics with radial velocity, $P\neq 0$, strong
singularities appear in more cases, as we show in the results comprised
in table \ref{table5}.

\begin{table}
\caption{\label{table5}Degree of singularity of timelike geodesics around
$t_{0}$.}
\begin{center}
    \begin{tabular}{llllll}
	\br
	${\eta_{0}}$ & ${\eta_{1}}$ & ${k}$ & $c_{0}$ &\textbf{Tipler} &
	\textbf{Kr\'olak}  \\ \mr
	$(-\infty,0)$ & $(\eta_{0},\infty)$ &  $0,\pm 1$ &$(0,\infty)$& 
	Strong & Strong  \\
	$0$ & $(0,1)$ &  && Weak & Strong  \\
	     & $[1,\infty)$ &   && Weak & Weak
\\
	$(0,1)$ & $(\eta_{0},\infty)$ &   && Strong & Strong  \\
	$1$ & $(1,\infty)$ & $0,1$  && Strong & Strong  \\
	 & $(1,\infty)$ & $-1$ &$(0,1)\cup(1,\infty)$ & Strong & Strong  \\
	 & $(1,3)$ &  & 1 & Weak & Strong  \\
	 & $[3,\infty)$ &  &  &Weak & Weak  \\
	$(1,\infty)$ & $(\eta_{0},\infty)$ & $0,\pm 1$  &$(0,\infty)$&
Strong & Strong \\
	\br
    \end{tabular}
    \end{center}
\end{table}

As we see, the results are essentially the same as for lightlike 
geodesics, with a difference of behaviour for models with coeficient 
$\eta_{0}$ lower or equal than minus one. Timelike geodesics do reach the 
cosmological milestone at $t_{0}$ in the form of a strong 
singularity. These models are lightlike geodesically complete, 
though they are timelike geodesically incomplete. In fact, table 
\ref{table5} provides the classification of FLRW cosmological 
models according to the strength of their singularities.

\section{\label{sec5}Conclusions}

In this paper we have provided a complete classification of FLRW 
cosmological models according to the strength of their singularities 
in terms of a generalised Puiseux expansion of the scale factor in 
coordinate time around cosmological milestones.

Though the velocity of the geodesic is finite at big rips, this does 
not prevent the appearance of strong singularities, except for two 
groups of models: those which behave at lowest order as Milne 
universe and those with sudden singularities. However, lightlike 
geodesics do not reach the big rip singularities for exponents 
$\eta_{0}$ lower or equal than minus one.

The results of Catto\"en and Visser in \cite{visser} are more 
restrictive, since they do not deal with the strength of the 
curvature singularities. For them models with $\eta_{0}=0$, $\eta_{1}\ge 2$
or $\eta_{1}=1$, $\eta_{2}\ge 2$, and those with $\eta_{0}=1$, $k=-1$,
$c_{0}=1$, $\eta_{1}\ge 3$ are free of polynomial scalar curvature 
singularities. These results coincide with ours for Milne-like models 
with Kr\'olak's definition for strong singularities, but include 
more models in the case of sudden singularities.

Their results for derivative curvature
singularities are even more restrictive. The only models which are 
free of such singularities are those with $\eta_{0}=0$ and
natural exponents $\eta_{i}$, $i\ge 1$ and those with $\eta_{0}=1$,
$k=-1$, $c_{0}=1$ and natural exponents $\eta_i\ge 3$, $i\ge 1$.

This is not surprising, since derivatives of the curvature tensor do 
not appear in our equations.

However, the main difference between both results arises from the 
fact that curvature singularities do not see that lightlike geodesics 
do not reach the singularity in finite proper time.

Though this classification of singularities conveys the idea of 
ubiquous singularities in FLRW cosmological models, it is worthwhile 
mentioning that singularities mostly appear in models with vanishing, 
divergent or non-smooth scale factors.

A similar scenario appeared in inhomogeneous scalar field Abelian
diagonal $G_{2}$ models \cite{wide}, where singularity-free 
cosmological models formed an open set. 

\ack

L.F.-J. is supported by the Spanish Ministry of Education and Science
Project FIS-2005-05198.  R.L. is supported by the University of the
Basque Country through research grant UPV00172.310-14456/2002 and by
the Spanish Ministry of Education and Culture through the RyC program,
and research grants FIS2004-01626 and FIS2005- 01181.

\end{document}